\documentclass{article}

\PassOptionsToPackage{numbers, compress}{natbib}

\usepackage[final]{neurips_mlsb_2021}

\usepackage[utf8]{inputenc} 
\usepackage[T1]{fontenc}    
\usepackage{hyperref}       
\usepackage{url}            
\usepackage{booktabs}       
\usepackage{amsfonts}       
\usepackage{nicefrac}       
\usepackage{microtype}      
\usepackage{xcolor}         
\usepackage{bm}             
\usepackage{graphicx}      
\usepackage{amsmath}        
\usepackage{listings}

\title{TERMinator: A Neural Framework for Structure-Based Protein Design using Tertiary Repeating Motifs}

\author{%
  Alex J. Li \\
  Department of Chemistry\\
  Massachusetts Institute of Technology\\
  \texttt{alexjli@mit.edu} \\
  \And
  Vikram Sundar \\
  Computational and Systems Biology Program \\
  Massachusetts Institute of Technology \\
  \texttt{vsundar@mit.edu} \\
  \And
  Gevorg Grigoryan \\
  Department of Computer Science\\
  Dartmouth College\\
  \texttt{gevorg.grigoryan@dartmouth.edu} \\
  \And
  Amy E. Keating\\
  Departments of Biology and Biological Engineering\\
  Massachusetts Institute of Technology\\
  \texttt{keating@mit.edu} \\
}

\begin{document}

\maketitle

\begin{abstract}
    Computational protein design has the potential to deliver novel molecular structures, binders, and catalysts for myriad applications. Recent neural graph-based models that use backbone coordinate-derived features show exceptional performance on native sequence recovery tasks and are promising frameworks for design. A statistical framework for modeling protein sequence landscapes using Tertiary Motifs (TERMs), compact units of recurring structure in proteins, has also demonstrated good performance on protein design tasks. In this work, we investigate the use of TERM-derived data as features in neural protein design frameworks. Our graph-based architecture, TERMinator,  incorporates TERM-based and coordinate-based information and outputs a Potts model over sequence space. TERMinator outperforms state-of-the-art models on native sequence recovery tasks, suggesting that utilizing TERM-based and coordinate-based features together is beneficial for protein design. 
\end{abstract}

\section{Introduction}

The goal of computational protein design is to identify a protein sequence that has a specified structure and, thereby, function \cite{Frappier2021}. Protein design has been used to design enzymes that catalyze important organic reactions \citep{Siegel2010} and to identify miniproteins that bind to the spike protein of SARS-CoV-2 and inhibit infection \citep{Cao2020}. However, protein design methods such as Rosetta, which are rooted in approximations of molecular physics, are too inaccurate to be reliably used without multiple rounds of expensive experimental trial-and-error \citep{Ingraham2019, Frappier2021}. Recently developed deep learning methods based on rigid coordinate featurizations have been applied to this problem and have achieved remarkable success using architectures like graph neural networks \citep{Ingraham2019} or 3D equivariant neural networks \citep{Jing2021}.


An alternative featurization of proteins using Tertiary Motifs (TERMs) has recently shown success quantifying the sequence-structure relationships needed for design using purely statistical methods \citep{Zhou2020}. TERMs are small, compact structural units that recur frequently in unrelated proteins and have been shown to cover a large portion of protein space \citep{Mackenzie2016}. A scoring function can be derived by defining TERMs within a given protein, searching for the closest matches across the PDB, and evaluating the sequence statistics of the resulting matches. TERMs are fuzzier than coordinate-based features, since they do not need to match exactly across different proteins.  Statistical models based on TERMs can capture sequence-structure relationships \citep{Zheng2015}, predict mutational changes in protein stability \citep{Zheng2017}, and be applied directly to protein design \citep{Zhou2020, Frappier2019}. 

The success of statistical models based on fuzzy TERMs suggests that these models capture information that can potentially augment models that consider only rigid coordinates. In this work, we designed a deep neural network that takes both TERM data and coordinate data as inputs and returns a scoring function that can be applied to evaluate any sequence on a given structure. Our method, named TERMinator, outperforms previous state-of-the-art methods on native sequence recovery tasks and we show through ablation studies that the use of TERM data is essential for the best performance. Our results suggest that TERMs provide a useful featurization of proteins for deep learning models.

\section{Methods}
    \subsection{Dataset}
        Following \citet{Ingraham2019} and \citet{Jing2021} we use the CATH 4.2 split curated by \citet{Ingraham2019}. For every chain, we describe the structure using a set of TERMs that includes sequence-local singleton TERMs (usually 3 contiguous residues) and sequence non-local pair TERMs (usually 2 interacting 3-residue segments). For each TERM, we perform substructure lookup \citep{Zhou2020} against a non-redundant database generated from the Protein Data Bank (PDB) on Jan. 22, 2019. Self-matches with the protein itself are discarded. For each match, we record the sequence along with a number of geometric features of the backbone. For more details about the TERM data, see the Appendix.
        
    \subsection{Architecture}
    \begin{figure}[t]
        \centering
        \includegraphics[width=\linewidth]{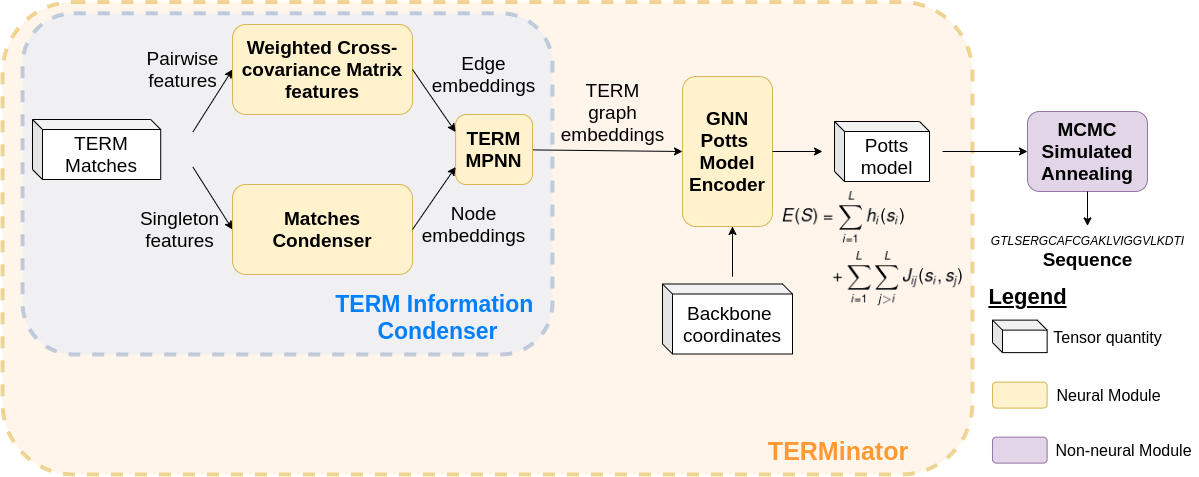}
        \caption{Model Architecture. The TERM Information Condenser extracts information from structural matches to TERMs in the target protein to construct node and edge embeddings. The GNN Potts Model Encoder takes in TERM data and coordinate features and outputs a Potts model over positional amino acid labels (see Appendix: Potts Model for functional form). We use MCMC simulated annealing to generate optimal sequences given the Potts model.}
        \label{fig:architecture}
    \end{figure}
    
        The network, shown in Figure \ref{fig:architecture}, can be broken into two sections. The first section, the TERM Information Condenser, learns local structure via graph convolutions on small, local TERM graphs. The second section, the GNN Potts Model Encoder, learns global structure via graph convolutions over nearest neighbors across the entire chain. Further details about the network architecture can be found in the Appendix.
        
        \paragraph{TERM Information Condenser}
            
            We represent TERMs as bidirectionally fully-connected graphs. Per TERM, we take the top 50 matches in the PDB with lowest RMSD. Each residue in each TERM match is converted to a set of geometric features describing backbone geometry and residue accessibility along with a one-hot encoding of the match residue identity. Node embeddings are generated from these initial features via the Matches Condenser, an attention-based pooling network that operates on the same residue across all matches per TERM. Edge embeddings, representing residue interactions within a TERM, are computed by constructing weighted cross-covariance matrices between the two residues' initial features, flattening them to vector form, and compressing them with a feedforward network.
            
            We feed these preliminary TERM graph embeddings through a 3-layer message-passing network known as the TERM MPNN, which operates on the local graph for each TERM. This results in a collection of residue embeddings and edge embeddings per TERM. We stitch together a full structure embedding by taking the mean of all replicate embeddings for both nodes and edges, since all residues and edges are covered by at least one TERM. 
            
        \paragraph{GNN Potts Model Encoder}
            The GNN Potts Model Encoder combines the structure embedding produced by the TERM Information Condenser and target protein backbone coordinate features to produce a Potts model over the structure's sequence space. We use the coordinate-based structure embedding presented in \citet{Ingraham2019} and concatenate the TERM-based structure embedding, where it overlaps with the global $k$-nearest neighbors graph of residues, to generate the initial full-chain representation (see the Appendix for further details). The GNN Potts Model Encoder is another message-passing network that is identical to the TERM MPNN in architecture, but instead operates on the global $k$-NN graph, including self-edges. We produce a Potts model from the output of this network by projecting each edge embeddings into a matrix of residue-pair interaction energies. Self-energies are defined as the diagonal of this matrix, while pair-energies are defined by the entire matrix.
            
    \subsection{Training}
        The loss function is the negative log \textit{composite psuedo-likelihood} averaged across residue pairs. The composite psuedo-likelihood is the probability that any pair of interacting residues has the same identity as that pair of residues in the target sequence, given the remainder of the target sequence. Unlike the classic psuedo-likelihood, the composite psuedo-likelihood depends on all values in the Potts model and thus penalizes pathological Potts models that only display reasonable energies on observed values. A mathematical definition of composite psuedo-likelihood in the context of a Potts model can be found in the Appendix. 
        
    \subsection{Performance Metrics}
    
    We evaluate performance by native sequence recovery, following \citet{Ingraham2019} and \citet{Jing2021}. For each Potts model, we sample $100$ sequences via MCMC-simulated annealing, reducing the temperature from $kT = 1$ to $kT = 0.1$. The lowest energy sequence is used to compute sequence recovery, and the sequence recovery for the entire test set is reported as the median recovery over all chains in the test set. We run all of our models in triplicate and report mean $\pm$ standard deviation.

\section{Results}

\subsection{Model performance and ablation studies}
    \begin{table}[ht]
      \caption{Ablation Studies of TERMinator. Native sequence recovery is listed as mean $\pm$ standard deviation for triplicate train/test runs on the same data split, where triplicate data are available.}
      \label{tab:results}
      \centering
      \begin{tabular}{ll}
        \toprule
        \textbf{Model version} & \textbf{Percent Sequence Recovery} \\
        \midrule
        TERM Information Condenser + GNN Potts Model Encoder & $41.73 \pm 0.27$  \\ 
        Ablate TERM Information Condenser & $40.29 \pm 0.26$ \\
        \quad Ablate TERM MPNN & $41.19 \pm 0.05$ \\
        \quad \textbf{Ablate Singleton Features} & $\textbf{42.22} \bm{\pm} \textbf{0.06}$ \\ 
        \quad Ablate Pairwise Features & $41.53 \pm 0.16$ \\
        Ablate GNN Potts Model Encoder & $29.94 \pm 1.10$\\
        \quad Ablate Coordinate-based Features, Retain $k$-NN graph & $35.87 \pm 0.18$\\
        GNN Potts Model Encoder Alone (no TERM information) & $39.66 \pm 0.30$      \\
        dTERMen & 24.32 \\ 
        \midrule 
        \citet{Jing2021} & 40.2\\ 
        \citet{Ingraham2019} (as reported by \citet{Jing2021}) & 37.3\\
        \bottomrule
      \end{tabular}
    \end{table}                                                     

    We train TERMinator, alongside several ablated versions, to understand how the different modules and input features affect performance on native sequence recovery. Results are shown in Table \ref{tab:results} and descriptions of the ablated models are in the Appendix. Our best models outperform state-of-the-art methods. Interestingly, our best performing model is TERMinator with the singleton features from the TERM matches ablated. This could be attributed to overfitting by the Matches Condenser, which uses an expressive attention-based mechanism on a relatively small number (50) of TERM matches. 
    
    Our ablation studies suggest that while TERM-based features and coordinate-based features are largely redundant, neither serves as a full replacement for the other. A TERMinator model trained purely on coordinate data can achieve a respectable sequence recovery of 39.7\%. Similarly, training on TERM data with no coordinate information (but with the  benefit of the global $k$-NN graph, an inherently fuzzy feature) also achieves a respectable recovery of 35.87\%. However, when TERMinator utilizes both source of data, it outperforms both models, as well as published state-of-the-art models, on native sequence recovery.

    We also note that when we ablate the GNN Potts Model Encoder, TERMinator effectively acts as a better version of dTERMen, a statistical framework which utilizes TERM information to compute Potts models over proteins \citep{Zhou2020}. The ablated TERMinator and dTERMen share access to essentially the same TERM input data (see the Appendix for a more detailed discussion), but the learned model makes better use of this information and is likely a better choice for future design tasks.

\subsection{Confusion Matrices}

\begin{figure}[ht]
    \centering
    \includegraphics[width=0.45\linewidth]{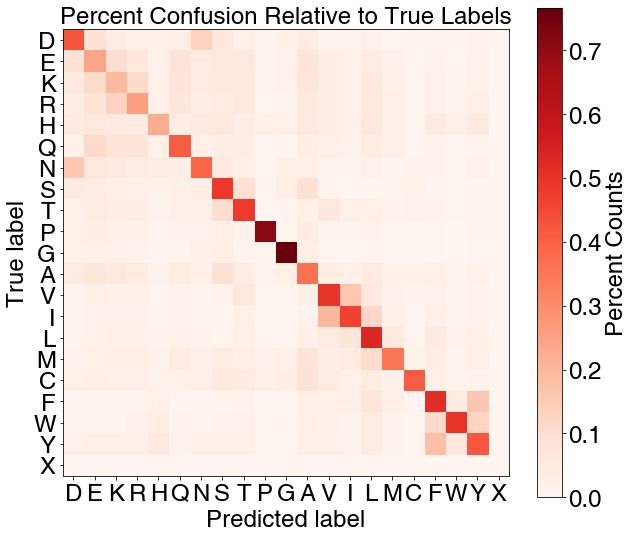}
    \caption{Confusion matrix reporting percent confusion with respect to true residue identity. Values report aggregate performance by the singleton-features-ablated TERMinator model across triplicate runs. `X' is an out-of-alphabet label used to represent non-canonical residues.}
    \label{fig:confusion_matrix}
\end{figure}

To better understand the performance of TERMinator, we examine the amino-acid confusion matrix. A representative matrix is shown in Figure \ref{fig:confusion_matrix}. The strong diagonal in the matrix reflects the high overall native sequence recovery; glycine (G) and proline (P) are particularly well recovered. Interestingly, the "mistakes" the model makes are physically realistic. We see confusion within the EKR block, which contains charged amino acids glutamate, lysine, and arginine, with the switch between charges potentially attributable to the model reversing the direction of salt bridges. Other blocks include: ST, with highly similar hydroxyl sidechains; VI, with sterically similar branched aliphatic sidechains; FWY, encompassing aromatic residues; and DN, with isosteric side chains. These substitutions are highly plausible and illustrate that TERMinator learns physicochemically realistic representations of proteins, lending confidence in the model's utility for design applications.


\section{Conclusion}

In this work, we present TERMinator, a graph-based network that utilizes both TERM and coordinate features to produce a Potts model over sequence space for any given protein chain. This model outperforms state-of-the-art neural models on the native sequence recovery task, and we attribute this performance boost to the inclusion of TERM data. Additionally, confusion matrices suggest that the model learns physically realistic representations. In future work, we hope to apply this model to real-world protein design tasks, as well as extend the model to be trainable over a variety of target outputs, such as binding affinity.

\begin{ack}
This work was funded by an award from the National Institutes of General Medical Sciences to G.G. and A.K., 5R01GM132117. The authors acknowledge Dartmouth Anthill, MGHPCC C3DDB, and MIT SuperCloud for providing high-performance computing resources used to generate research results for this paper. A.L. acknowledges funding from MIT UROP and would like to thank Sebastian Swanson for teaching him how to use the dTERMen software suite. V.S. acknowledges funding from the Fannie and John Hertz Foundation. 
\end{ack}

\newpage

\section*{Appendix}
\subsection*{Potts Model}
Our model outputs a Potts model over positional amino acid labels,  commonly known as an energy table in the protein design community. A Potts model describes a mapping from sequence $S$ of length $L$ to energy $E(S)$ with the functional form
$$ E(S) = \sum_{i=1}^L h_i(s_i) + \sum_{i=1}^L \sum_{j>i}^L J_{ij}(s_i, s_j) $$
where 
\begin{itemize}
    \item Singleton terms $h_i(s_i)$ describe the energy contribution of position $i$ in $S$.
    \item Pairwise interaction terms $J_{ij}(s_i, s_j)$ describe the energy contribution from the interaction between positions $i$ and $j$ in $S$.
\end{itemize}

In our Potts model, the singleton term takes the form $h_i(s_i) = E_s(R_i = m)$, where $E_s$ is a lookup table of energies for placing residue $m$ at position $i$. The pairwise interaction term takes the form $J_{ij}(s_i, s_j) = E_p(R_i=m, R_j=n)$, where $E_p$ is a lookup table of energies of placing residue $m$ at position $i$ and residue $n$ at position $j$. This functional form is attractive because it can be used to rapidly evaluate any sequence's energy, and is easy to optimize via MCMC-based methods.

\subsection*{TERM data}
Our goal was to feed into our neural network similar data as would normally be mined by dTERMen \citep{Zhou2020}. This would enable us to differentiate the limitations of TERM data themselves from the limitations associated with the specific statistical approach in dTERMen. To this end, we modified an in-house version of the dTERMen program with the ability to output TERM match information for all of the motifs used in the standard procedure (as described in \citet{Zhou2020}). Briefly, dTERMen defines three types of TERMs in the input structural template: singleton, near-backbone, and pair TERMs. Singleton TERMs are defined around each residue $i$ via the contiguous fragment between residues $(i-n)$ and $(i+n)$, where $n$ is a parameter ($n=1$ was used in this study). Near-backbone TERMs combine the local backbone around residue $i$ (i.e., the singleton fragment) with local backbone fragments around each residue $j$ whose backbone is geometrically poised to interfere with amino-acid sidechains at $i$. Finally, pair TERMs are defined around each pair of residues $i$ and $j$ that are geometrically poised to affect each other's amino-acid choice.

As described in Zhou et al. \citep{Zhou2020}, it is frequently the case that the full near-backbone TERM around a residue (i.e., the generally multi-segment motif that captures all relevant surrounding backbone fragments) does not contain sufficient structural matches in the database, in which case the dTERMen procedure seeks to optimally partition the overall near-backbone contribution into as few sub-motifs as possible. This step adds considerable search time. We reasoned that a learning-based approach may be better at extracting relevant statistical couplings between residue sites, such that a detailed breakdown of sidechain-to-sidechain versus backbone-to-sidechain coupling statistics may not be necessary. Therefore, we omitted near-backbone TERMs in this study for computational efficiency.

In finding close structural matches, dTERMen uses a motif complexity-based empirical RMSD cutoff (defined in Mackenzie et al. \citep{Mackenzie2016}), with additional settings used to control the minimal and maximal number of matches. In this study, we set these limits to lower values than previously reported \cite{Zhou2020} for computational efficiency. Specifically, the minimal/maximal match counts were 200/500 for singleton TERMs, and 400/500 for pair TERMs. Under these settings, dTERMen takes roughly 4 minutes per residue (single-core, 8GB RAM). The native sequence recovery rate of dTERMen was estimated on the basis of energy tables produced with these settings. As input into the neural-network models, only data from the top 50 TERM matches were used.

\begin{figure}[ht]
    \centering
    \fbox{\begin{minipage}{0.9\linewidth}
    \fontsize{9pt}{9pt}\selectfont
    \texttt{
* TERM k \\
list of position indices covered by the TERM (suppose N of them)\\
sequence of match 0; RMSD; N phi values; N psi values; N environment values\\
sequence of match 1; RMSD; N phi values; N psi values; N environment values
}
\begin{center}
    \texttt{...}
\end{center}
    \end{minipage}}
    \caption{TERM matches information structure.}
    \label{fig:term_data}
\end{figure}
For each considered TERM, we output which positions of the structural template it covers along with information on each of its matches in the database of known structures, i.e. the match sequence, best-fit backbone RMSD from the query, backbone $\phi$ and $\psi$ values at each residue, and the "environment" of each residue--a scalar ranging from 0 to 1 that describes how solvent-exposed the residue is (the freedom metric defined in \citep{Zheng2017}); see Figure \ref{fig:term_data}.

Additionally, for every residue in a TERM we compute a ``contact index'' that specifies the sequence distance from a central residue in the TERM. TERMs are constructed either around a single central residue (for a singleton TERM) or a pair of residues (for a pair TERM). We assign central residues an index of 0 and define the contact index for the remaining residues as the directional sequence distance to the closest intra-chain center residue. More specifically, non-central residues closer to the N-terminus than their corresponding central residue are assigned a negative integer contact index, while non-central residues closer to the C-terminus than their corresponding central residue are assigned a positive integer contact index.

\subsection*{Matches Condenser}
The Matches Condenser serves to compress the per-residue matches information across all TERM matches into one latent residue representation. The initial featurization of each residue per TERM match consists of:
\begin{itemize}
    \item a one-hot encoding of the match residue's identity
    \item the residue's torsion angles lifted to the $3$-torus $\{\sin, \cos \} \times \{\phi, \psi, \omega\}$
    \item the residue's environment value
    \item the TERM match's overall RMSD.
\end{itemize}
         
The Matches Condenser operates using an attention-based pooling mechanism, reminiscent of the CLS token used in BERT \citep{Devlin2018}. Along with every set of TERM matches, we generate and feed in a ``pool token'' per residue as an additional match. The intuition behind the pool token is that the Matches Condenser is able to update the pool token with the most important summary information from the TERM matches through multiple rounds of self-attention updates. The pool token is intended to capture the most important information about TERM matches, which we can then take as node embeddings for the TERM graph. 

For every residue in the TERM, we concatenate the residue's true torsion angles (lifted to the 3-torus) and its environment value, and feed this vector through a linear layer to create a pool token. We also create an associated set of ``target vectors'', which are derived from the same set of features as the pool tokens but are instead used in the attention mechanism itself. First, the base features per TERM match are fed through a two layer dense network. Then, across all TERM matches as well as the pool token, we perform MatchAttention, four rounds of alternating multi-headed self-attention ($n=4$) and feedfoward updates. For the attention computation, key and values vectors are computed by concatenating the target vector and the current match embedding and projecting using a linear layer, while query vectors are computed solely by projecting the current match embedding using a linear layer. In other words, for residue $n$ in TERM $t$, with the $i$th match embedding $h_{t,n,i}$ and target structure information embedding $\tau_{t,n}$, we compute query $q$, key $k$, and value $v$ as
\begin{align*}
    q &= W_Q [h_{t,n,i}]\\
    k &= W_K [h_{t,n,i}; \tau_{t,n}]\\
    v &= W_V [h_{t,n,i}; \tau_{t,n}]\\
\end{align*}
where $W_Q, W_K, W_V$ are linear layers and $[;]$ is the concatenation operator.

\subsection*{Weighted Cross-Covariance Matrix Features}
         
For each pair of residues, we compute the weighted cross-covariance matrix between the initial featurization of the TERMs across all matches. Let $r_i$ represent the RMSD of match $i$. Then, the weight of a match is computed as
$$w_i = \frac{e^{-r_i}}{\sum_{j} e^{-r_j}}$$
This matrix is then flattened into a vector and  its dimensionality is reduced using a two-layer feedforward network with ReLU activations. This output is used as the edge feature between the two residues of concern in the TERM graph.

\subsection*{TERM MPNN}
        
We define the following notation:
\begin{itemize}
    \item $h_{i,t}$: the embedding for residue $i$ in TERM $t$
    \item $h_{i \rightarrow j, t}$: the embedding for directional edge $i \rightarrow j$ in TERM $t$
    \item $f_n, f_e$: three-layer feedforward networks with ReLU activations
    \item $g_n, g_e$: two-layer feedfoward networks with ReLU activations
    \item $\mathcal{N}_t$: the set of residues in TERM $t$
    \item $s_{i,t}$: the sinusoidal embedding of the contact index of residue $i$ in TERM $t$ (see positional embedding in \citet{Vaswani2017})
    \item $[;]$ represents the concatenation operation
\end{itemize}

The TERM MPNN utilizes alternative edge-update and node-update layers. The update for a directional edge $i \rightarrow j$ in TERM $t$ is computed as
\begin{align*}
    h_{i \rightarrow j,t,\text{update}} = \frac{1}{2} \Big[ f_e([h_{i,t}; s_{i,t}; h_{i\rightarrow j, t}; h_{j,t}; s_{j,t}]) + f_e([h_{j,t}; s_{j,t}; h_{j\rightarrow i, t}; h_{i,t}; s_{i,t}]) \Big]
\end{align*}
And this update is applied as follows:
\begin{align*}
    h_{i\rightarrow j, t} &\leftarrow \text{LayerNorm}(h_{i \rightarrow j,t} + \text{Dropout}(h_{i \rightarrow j,t,\text{update}}))\\
    h_{i\rightarrow j, t} &\leftarrow \text{LayerNorm}(h_{i \rightarrow j,t} + \text{Dropout}(g_e(h_{i \rightarrow j,t})))
\end{align*}

The update for a node is computed as
\begin{align*}
    h_{i,t,\text{update}} = \frac{1}{\left| \mathcal{N}_t \right|} \sum_{j \in \mathcal{N}_t} f_n([h_{i,t}; s_{i,t}; h_{i\rightarrow j, t}; h_{j,t}; s_{j,t}])
\end{align*}
And this update is applied as follows:
\begin{align*}
    h_{i, t} &\leftarrow \text{LayerNorm}(h_{i,t} + \text{Dropout}(h_{i,t,\text{update}}))\\
    h_{i, t} &\leftarrow \text{LayerNorm}(h_{i,t} + \text{Dropout}(g_e(h_{i,t})))
\end{align*}

The TERM MPNN contains three layers, with each layer containing an edge update followed by a node update. After these updates, all bidirectional edges are merged into undirected edges via taking the mean of the two edge embeddings. 

\subsection*{GNN Potts Model Encoder}

The GNN Potts Model Encoder is another message-passing network that is identical to the TERM MPNN in architecture but takes in different input features. The GNN Potts Model Encoder operates on a $k$-NN graph rather than a fully-connected graph, meaning node updates are computed over a residue's $k$ nearest neighbors. Additionally, because there is no notion of ``contact index'' when it comes to global structure, the update function does not take such features as inputs. 

Before running message passing, the GNN must first stitch together the TERM-based structure embedding and the coordinate-based structure embeddings. Node embeddings for the GNN are computed by concatenating the coordinate-based features from \citet{Ingraham2019} and the TERM-based features and feeding that vector through a linear layer to compress the vector back to the original dimensionality. Edge embeddings are also formed by computing the coordinate-based edge embedding from \citet{Ingraham2019}, concatenating the corresponding TERM edge embeddings, and feeding that vector through a linear layer to compress the vector back to the original dimensionality. In the case that a TERM edge embedding does not exist for that particular $k$-NN graph edge, a zero-vector of equal dimensionality is used instead.

The edge embeddings derived after message-passing is completed are then projected to a $400$-dimensional vector by a feedforward network and reshaped to form a matrix containing interaction energies between pairs of interacting residues. The interaction energy matrices give the pair energies of the Potts model. Due to the inclusion of self-edges in the $k$-NN graph, we can also compute self-energies for the Potts model by taking the diagonal of the self-interaction matrix produced by the self-edge for each residue.

\begin{figure}[t]
    \centering
    \includegraphics[width=\linewidth]{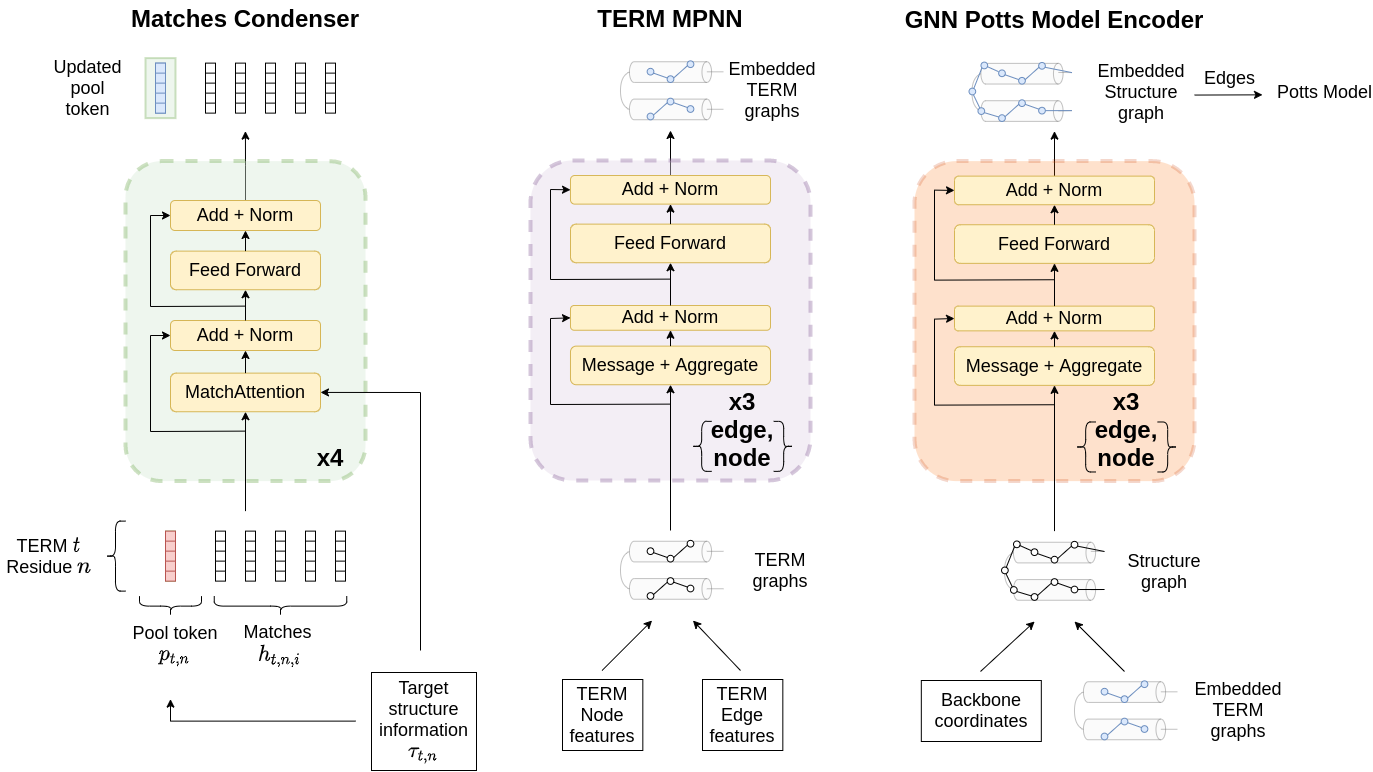}
    \caption{TERMinator Submodule Architectures.}
    \label{fig:arch_detas}
\end{figure}

\subsection*{Loss Function and Training}
Our loss function is derived from a quantity known as ``composite psuedo-likelihood'', the probability that any pair of interacting residues has the same identity as that pair of residues in the target sequence, given the remainder of the target sequence. Composite psuedo-likelihood can be defined using the energies described by the Potts model as follows. As stated in Appendix: Potts Model, a Potts model is defined by two functions: the self-energy function $E_s(R_i = m)$ evaluates the energy of residue $i$ with identity $m$, and the pair-energy function $E_p(R_i = m, R_j = n)$ evaluates the energy of residue $i$ with identity $m$ interacting with residue $j$ with identity $n$. From the Potts model, we compute the contextual pairwise energy $E_{cp}$ for a sequence with residue $i$ having identity $m$ and residue $j$ having identity $n$, provided all other residues $R_u$ with identity $r_u$, as:
\begin{align*}
    &E_{cp}(R_i=m, R_j=n, \{R_u = r_u\}) =  \\
                        &\quad\quad E_s(R_i=m) + E_s(R_j=n)\\
                        &\quad\quad + E_p(R_i=m, R_j=n)\\
                        &\quad\quad + \sum_{u \neq m,n} \left( E_p(R_i=m, R_u=r_u) + E_p(R_u = r_u, R_j=n) \right)\\
\end{align*}
From this energy, we can compute the composite psuedolikelihood $p(R_i = m, R_j = n, \{R_u = r_u\})$:
$$ p(R_i = m, R_j = n, \{R_u = r_u\}) = \frac{\exp[-E_{cp}(R_i=m, R_j=n, \{R_u = r_u\})]}{\sum_{k,l} \exp[-E_{cp}(R_i=k, R_j=l, \{R_u = r_u\})]}$$

We train TERMinator to minimize negative log composite psuedo-likelihood, averaged across all pairs of interacting residues. Training occurs over 100 epochs using the Noam learning rate scheduler \citep{Vaswani2017}. Training time ranges from 3-5 days, depending on the particular ablations of TERMinator (using 2 NVIDIA Tesla V100 GPUs). Inference time on the Ingraham test set was on average 97 ms, or 61 $\mu$s/residue (using 1 NVIDIA Tesla V100 GPU).

\subsection*{Hyperparameters}
TERMinator uses two different hidden dimensionalities: the TERM Information Condenser uses a hidden dimension of 32, while the GNN Potts Model Encoder uses a hidden dimension of 128. This is largely due to GPU memory issues, as raw TERM data are much larger than coordinate data. Given more compute power, one direction to explore is how the model's performance is affected by varying the hidden dimensionality of both portions of the network.

\subsection*{Description of Ablated Models}

The following list provides a brief description of TERMinator ablation models:
\begin{itemize}
    \item \textbf{TERM Information Condenser + GNN Potts Model Encoder:} All neural modules included.
    \item \textbf{Ablate TERM Information Condenser:} The TERM Information Condenser is reduced to a series of linear transformations.
    \begin{itemize}
        \item \textbf{Ablate TERM MPNN:} Singleton and pairwise TERM features are directly passed to the GNN Potts Model Encoder.
        \item \textbf{Ablate Singleton Features:} Singleton TERM features are set to 0.
        \item \textbf{Ablate Pairwise Features:} Pairwise TERM features are set to 0.
    \end{itemize}
    \item \textbf{Ablate GNN Potts Model Encoder:} Outputs of the TERM Information Condenser are embedded on a $k$-NN graph and then projected to form a Potts Model.
    \begin{itemize}
        \item \textbf{Ablate Coordinate-based Features, Retain $k$-NN graph:} Coordinate-based features are set to 0. The $k$-NN graph is still retained.
    \end{itemize}
    \item \textbf{GNN Potts Model Encoder Alone (no TERM information):} Outputs of the TERM Information Condenser are set to 0.

\end{itemize}   
\textit{Ablate GNN Potts Model Encoder vs. dTERMen}

In the main text, we claim that the version of TERMinator with the GNN Potts Model Encoder ablated has access to essentially the same features as dTERMen. In order to make a fair comparison, it is important to acknowledge a few slight differences in the precise inputs. Regarding TERMinator, while this particular ablation form does not have access to coordinates directly, it does have access to a $k$-NN graph, which dTERMen does not get. However, when we ablate the GNN Potts Model Encoder, it has no opportunities to perform graph operations over the $k$-NN graph; instead, the $k$-NN graph is only used to restrict pair interactions in the Potts model to those present in the graph. Due to the nature of TERMs being constructed out of small sets of spatially-proximal residues (<7 residues), it is almost always the case that all TERM edges will be included in the $k$-NN graph (in this work, $k=30$), leading to negligible utilization of the $k$-NN graph itself. On the other hand, the version of dTERMen used uses more matches to compute the Potts model than does TERMinator, which was restricted to using the top 50 matches. We also note that the published version of dTERMen uses near-backbone TERMs that were not included for the dTERMen sequence recovery results reported here, although this also means that TERMinator did not have access to these TERMs either. All things considered, it is reasonable to assume that this ablation of TERMinator and dTERMen effectively have access to the same types of information, with dTERMen performing worse despite having access to more matches.
\newpage

\bibliographystyle{plainnat}
\bibliography{TERMinator}

\end{document}